# HYBRID DESIGN TOOLS - MAKING OF A DIGITALLY AUGMENTED BLACKBOARD.

**PhD. Lect. Arch. Camil Octavian Milincu** [1]
**PhD. Arch. Otilia Alexandra Tudoran** [1]
**Eng. Paul-Florin Tarce** [2]
**PhD. Lect. Eng. Ovidiu Banias** [2]

[1] The Politehnica University of Timisoara Faculty of Arhitecture and Urbanism, Architecture Dept., **Romania**
2 The Politehnica University of Timisoara Faculty of Automation and Computer Science, Automation and Applied Informatics Dept., **Romania**

## ABSTRACT

The way that design is being taught is continuously changing under the pressure of the transition from analogical to digital environments. This becomes even more important as the novelty and the alleged superiority of the digital world is used as a marketing tool by competing universities. Even though in some fields of application this approach is desirable, some particular aspects of teaching design and architecture make this transition debatable. The advantages of drawing on blackboards over drawing on whiteboard surfaces in regards of line aesthetic and expression possibilities were previously identified, along with the complementary necessary features for improvement. This study showcases a proof of concept in digitally augmenting a blackboard surface. The system allows the capturing, processing and making real time projections of images over the blackboard surface as trace references. Such a hybrid system, along with providing support for design and architecture related presentations and discussions could also mediate the contradictory relation towards technology that students and teachers have.

## INTRODUCTION

At the moment we are witnessing a change regarding the student's profile in terms of their relationship with technology. If until recently, as did the teachers, they were from the category of digital immigrants, nowadays students can be considered digital natives. Although they had access to digital tools and the Internet, students are not familiar to 3D design and modeling tools and they have to accumulate the necessary skills along the way. Previous studies [1] show the limitations that occur in the design process when trying to use exclusively digital tools without having an advanced level of competence. Even worse, there is a reluctance on the part of students to use analogue media in spite of the clear benefits it offeres in certain situations . This inefficient method is assumed, most of the students considering the use of anything other than digital means as a sign of weakness.

This study proposes the use of a system that aims to support the idea of working in a hybrid system, using discerningly any means at disposal, the sole purpose being the rapid development, and with minimal ingrades of the design solutions. The advantages of such a method have previously been identified by several studies [2], [3], [4], [5]. Thus, it is proposed to implement a hybrid system composed of a blackboard augmented with digital elements to be used in the presentations made during the courses and during the discussions during the design workshop. The choice of a blackboard in place of the whiteboard was made after a comparative study of the surfaces, used according to the needs of the design workshops. [6].

Similar studies in the field are based on interactive whiteboard systems (IBW). Student's interest and engagement may increase, even if user adaptation is needed [7]. Although in some areas these have clear advantages [8], in the case of architecture and interior design they are cumbersome to use and have a number of disadvantages. There are opinions that IBW systems have the drawback of continuing to focus on traditional teacher pedagogy elements. [9], [10]. In this present case this is not an issue, the system being used in the part of teacher presentations and joint discussions, the rest of the design classes being focused on the individual solutions of the students.

**PROPOSED HARDWARE & SOFTWARE SOLUTION**

The proposed system consists of 3 subsystems: desktop (PC) application with the photo camera, mobile application and microcontroller application. The ultimate goal is to capture photos using a camera or video camera connected to the PC, editing them and displaying them on the blackboard using a video projector. The desktop application can be used for the most part by using the keyboard for taking pictures, displaying, saving or deleting them, zooming in/out, rotating them, changing the contrast or brightness, and cropping the photos (using the mouse). In order to make this application easier to use and not having to move to the laptop or PC during the presentation, a mobile app that integrates the most important commands and functionality of the application has been built. The mobile app will communicate with the desktop application via Bluetooth. The mobile application allows direct and remote camera control, calibration, image processing and real-time display on the projector. While presenting lessons or projects, the user can scroll through the pictures made or imported, using the application as a slideshow. The user can also easily switch between viewing files and photos taken with the application. Figure 1 shows the block diagram of the system underlying this project. It describes the three subsystems that build the entire application (Android mobile application, Desktop PC application and application on the Arduino microcontroller). The central entity in this system is the Desktop-PC application, which is basically the core of this system. The Desktop application is also the most complex application and is directly or indirectly (in the case of the mobile application) linked to the other subsystems and elements in this project. The user can use this system as shown in the figure directly and indirectly. The user can directly use the PC on which the Desktop application is installed. In this mode, the other two applications (mobile and Arduino) may not be used, without this affecting the proper functioning of the application in general. Indirectly, the user can use some of the desktop application's features on the channel: mobile app - bluetooth - Arduino - usb - desktop application. The mobile app can send some basic commands without requiring permanent presence on the laptop / PC, making the system easier to use .

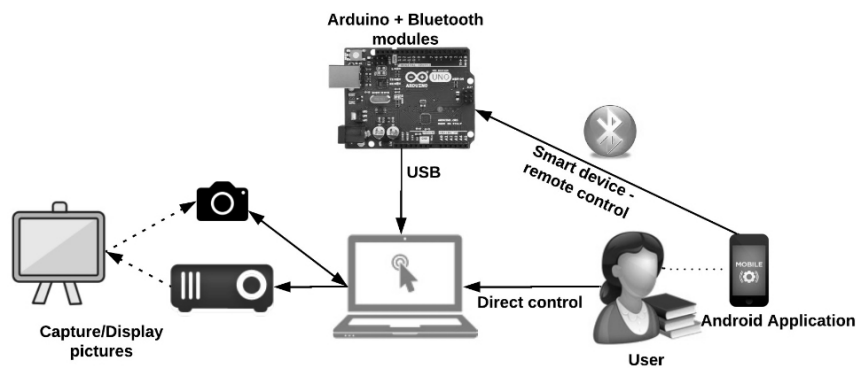

Fig. 1 System architecture

On the left side of Figure 1 are the capture and display/projecting devices. The displays consist of a video projector or a monitor/screen, which are interpreted by the application as secondary screens. The connection between the laptop / PC and these devices may be HDMI, RS232 or other connection. As a place to display the photos, which is not a feature of the application, it is a whiteboard (as specified in chapter 1 of the paper), but also other surfaces (walls, projection screens, etc.). A camera, a camcorder, or a webcam connected to your PC is used to take pictures. The capture camera can be a photo (tether shooting), video or webcam (capture frames capture). To be easy and fast to use, many of the application controls can be made from keys. For more advanced features such as cropping, changing the contrast and brightness, saving and sharing photos by Email, one needs to use both the mouse and the keyboard. However,in most cases, the functionalities needed to support a course/laboratory/seminar can be accessed via keyboard buttons. To be able to remotely control this system, which can be useful, it can connect to the PC through the Arduino (Bluetooth) module. Arduino will download Bluetooth data from the Android mobile app.The Android mobile app integrates commands for the most important features of this system, generally those that can also be accessed by using PC keys.

Characteristics of system components as shown in Figure 2: PC: Lenovo Essential B5400, Intel Core i5-4200M 2.50GHz, 4GB RAM, 500GB Hdd, nVidia Geforce 820M 2GB, Video Camera: Logitech C922 Pro Stream Webcam: Full HD 1080P with 30FPS and 720P with 60FPS, Arduino Uno Module (+ wires / connectors, 1kΩ resistors, 2kΩ), Bluetooth Module HC-05, Projector, Mobile Phone Samsung Galaxy S3 Neo

The desktop app was built using Microsoft's .NET Framework technologies. The mobile application was developed in Java. The Android Studio IDE development environment has been used, and it is an application for the Android operating system. Also, the mobile application by sending remote commands (wireless) needed a Bluetooth socket. The data is transmitted via the Bluetooth protocol to the Bluetooth module HC-05 that is connected to the Arduino Uno microcontroller. These two components were programmed into the Arduino IDE development environment using C ++. The data transmission protocol from the Arduino + Bluetooth HC-05 module to the PC is USB, the Arduino module being connected to a USB TypeB cable to it.

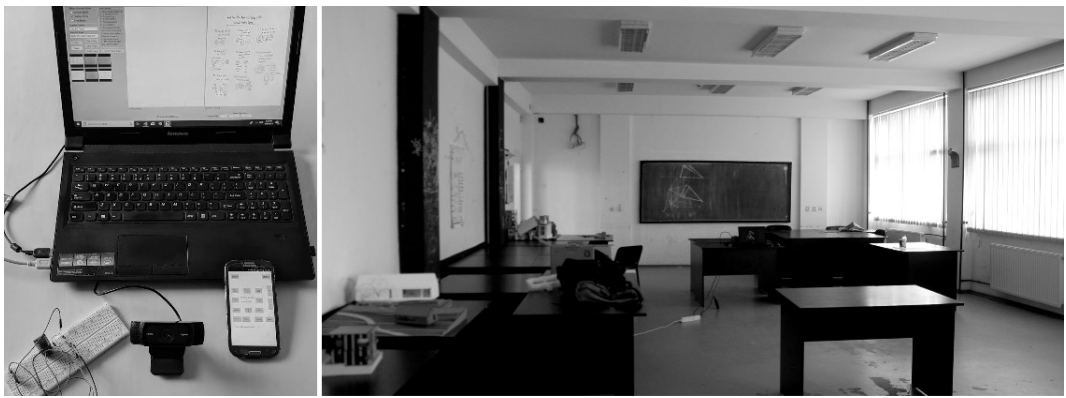

Fig. 2 Sistem components and classroom environment

## SYSTEM IN USE

In order to facilitate interaction and to minimize the time required for adaptation, a minimum number of tools were identified, in order to have the greatest degree of adaptability to the specific requirements of the workshops.

1. Dot Grid reference used for improving the graphic quality of drawings. The grid allows the display of an orthogonal or rotated reference grid, correlated with a graphical scale. Thus, one of the important problems of drawing on the board, namely the deformation of the drawings or the deviation from the orthogonal system is diminished, especially when the drawings become large enough to require the users's changing of position. In addition, a reference for maintaining a uniform system of units is provided. There is a possibility for making drawings to scale, drawn freehand, without the use of specialized tools which are always avoided, being cumbersome. This also allows for accurate and fast explorations in the initial project phases. Besides the fact that the drawings are more accurate, the system can encourage teachers to draw more on the board, providing personalized responses to the raised problems without the need for preparing digital slides.

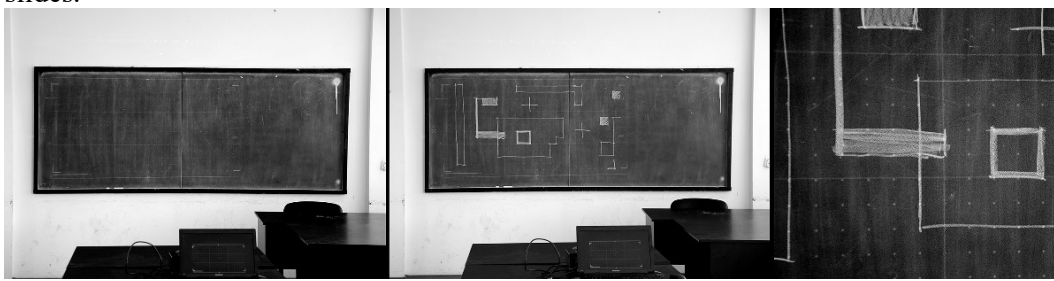

Fig.3. Grid tool: Dot grid, drawing and detail

2. Providing references for ergonomic studies or rough dimensioning of various components. Unlike common presentations, the system allows for punctual adaptation to the requirements specific to each course or theme presented. The information can be presented gradually, it can be hierarchized, focusing on some components. Equally important, possible sources of frequent errors in design can be highlighted. In the case of furniture design, it is possible to use trace references of imposed ergonomic dimensions, thus limiting the exploration of non-feasible variants.

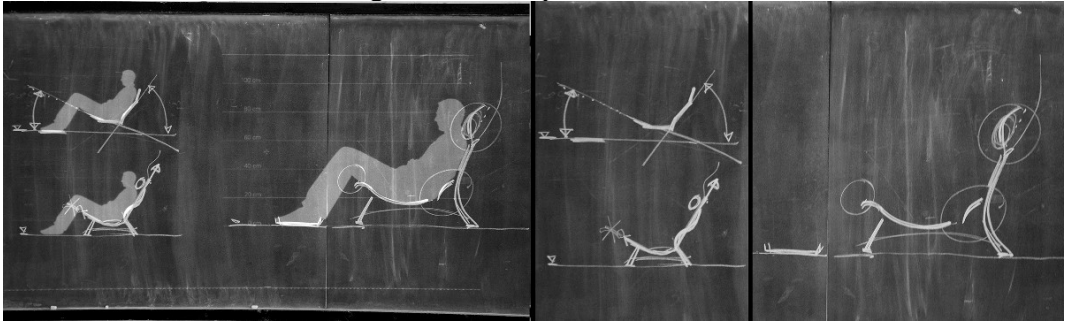

Fig. 4 Trace references for furniture design

3. Preset references for complex drawings. The tool is useful whenever complex geometric representations are required for descriptive geometry and related classes. It makes it possible to explain step-by-step drawing methods, without the need for cumbersome drawing aids. Using a reference is important especially because it does not allow the perpetuation of errors in the manual drawing that would make impossible for the teacher to complete the demonstration. In this case, the advantage of using a blackboard is particularly remarkable, as it is possible to generate lines with different intensities and thicknesses using only a piece of chalk

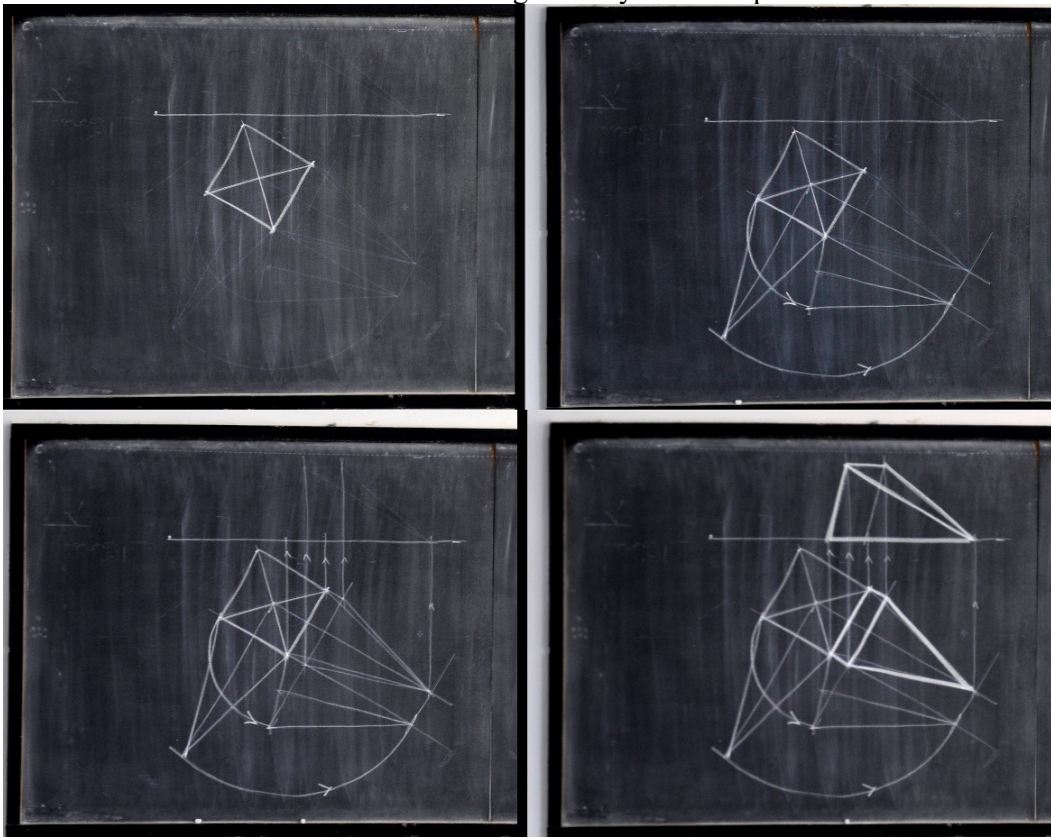

Fig. 5 Complex freehand drawing made using a trace reference

4. Recalling a previous image, captured from the blackboard. This can be used to continue or modify a previous presentation, which is very difficult to do after erasing the blackboard. As can be seen in fig. 6, there is no significant difference between the original image taken from the board and the projected image. Artifacts of the projector's matrix are only visible from a small distance away from the blackboard

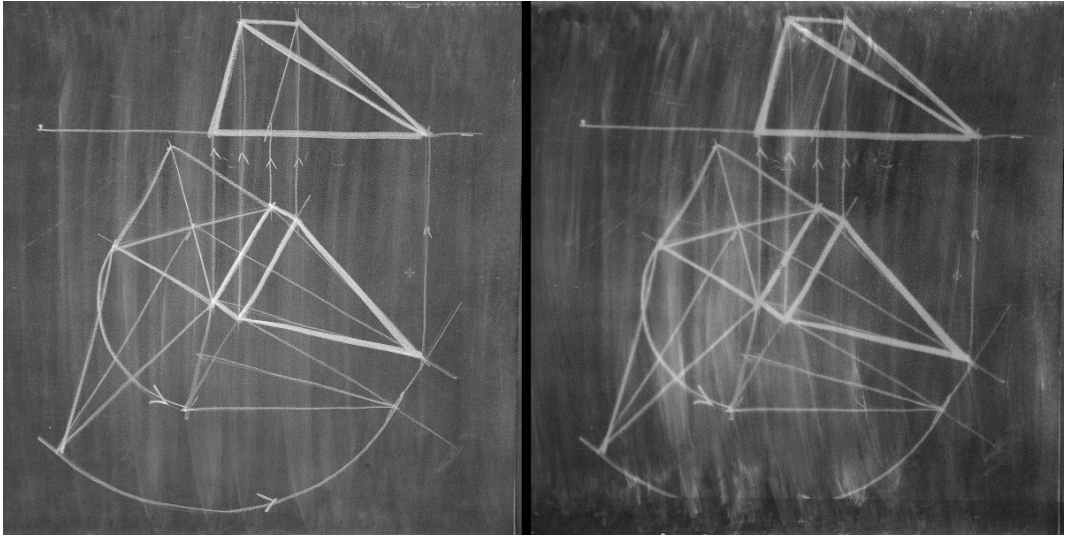

Fig. 6 Original chalk drawing and projected capture image on erased blackboard

5. It is possible to continue or modify previous presentations using projected references, without the need to redraw the entire drawing. The modifications can be made in real time, according to the student' feedback, without the need for prior prepared slides.

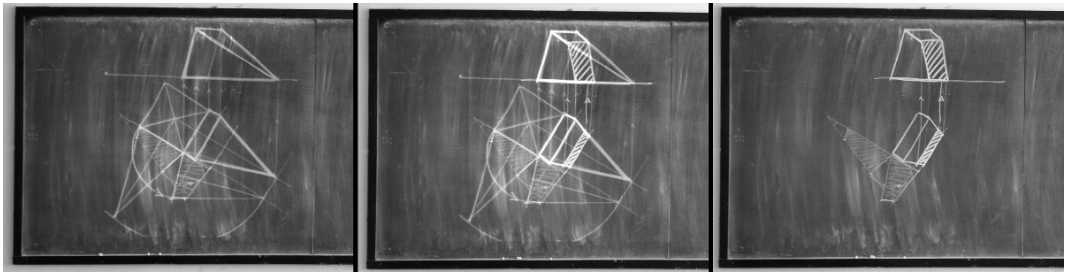

Fig. 7 Projected reference, continued drawing with reference, chalk drawing

6. The reference may consist of an image of the existing situation in which a building or interior space to be arranged or furnished is inserted. This preserves the proportions of the built environment without the need for 3D modeling of the context. For the study phase of possible approaches, 3D modeling of the context in which the project is situated often requires lengthy times due to its volume or complexity. In addition, working this way places emphasis on plausible viewpoints from which the subject is perceived to the detriment of inaccessible points of view made in order to accentuate some concepts.

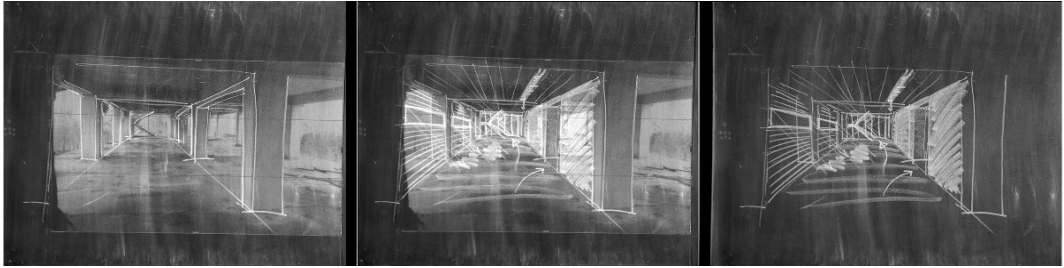

Fig. 8 Sketches made using a picture reference

**DISCUSSION. OTHER SYSTEM BENEFITS:**

This type of presentation forces the presenters to be active. Running a slideshow often becomes monotonous, the presenter having a fixed position in the hall, his role being reduced to the presentation and rendering of the images. The students' disconnection phenomenon can also be favored by knowing that the presentation is a preset one, and it can also be obtained as an appendix to course notes. On the other hand, a presentation using a digital augmented table can provide answers to point questions, being more adaptable to a preset digital presentation. In addition the multisensory perception should not be ignored either. The tactile and auditive [11] feedback is as important as gesture [12] and can bring new elements in a conversation. In the case of whiteboards, these have a smaller share, tactile and auditive elements becoming important when using a blackboard. In this case, there is a strong tactile correlation between the pressure and the speed of drawing given by the chalk, just like an important auditive component.

Using a blackboard leaves visible traces. Unlike the exclusive use of a projector, the use of the blackboard may have an impact on the perception of the students following this specialization. This becomes visible when classrooms are changed by the students. There is a difference between entering a classroom with a neutral setting, dominated by the white projection screen, and a classroom with a central element that still bears the traces of a presentation that has just ended.

**CONCLUSION:**

To avoid influencing students when a feed-back is requested, even anonymously, it is suggested that system verification be done by tracking over time, during a school year, how a hybrid workflow in the project framework becomes accepted. Although at the time of evaluations, when problems due to time management were obvious, students are aware of the fact that the full digital work method is not efficient, the hybrid mode is viewed with reluctance, the possible source being peer pressure from colleagues.

A hybrid mode of work has several advantages. It shortens the time spent on drafting, allowing a more extensive exploration of the solution. It can help avoiding the limitation of the vocabulary to what the student can easily mold into the design software. The problem is further aggravated by the tendency to use inventory items found in various libraries.

Another aspect is the environment in which the architectural, interior and furniture projects unfold. The works take place in the real, imperfect built environment. In addition, unpredictable issues that require a quick response from the designer occur within the jobsite. There is no time for making complex digital presentations.

In addition, it is necessary to familiarize yourself with the real environment in time, where perfect orthogonal systems are not found and where acceptance and provision of tolerances are necessary for a work to be accomplished successfully. The sterile digital environment leads to disconnection, it is necessary that the real environment imperfections be accepted and assumed.